# Sumanene monolayer of pure carbon: a two-dimensional Kagome-analogy lattice with desirable band gap, ultrahigh carrier mobility and strong exciton binding energy

Xiaoran Shi[1], Weiwei Gao[1], Hongsheng Liu[1], Zhen-Guo Fu[2*], Gang Zhang[3], Yong-Wei Zhang[3], Junfeng Gao[1*], Jijun Zhao[1]

1. *Key Laboratory of Materials Modification by Laser, Ion and Electron Beams, Dalian University of Technology, Ministry of Education, Dalian 116024, China*

2. *Institute of Applied Physics and Computational Mathematics, Beijing 100088, China*

3. *Institute of High Performance Computing, A*STAR, 138632 Singapore*

**ABSTRACT:** Design and synthesis of novel two-dimensional (2D) materials that possess robust structural stability and unusual physical properties may open up enormous opportunities for device and engineering applications. Herein we propose a 2D sumanene lattice that be regarded as a derivative of the conventional Kagome lattice. Our tight-binding analysis demonstrates sumanene lattice contains two sets of Dirac cones and two sets of flat bands near the Fermi surface, distinctively different from the Kagome lattice. Using first-principles calculations, we theoretically suggest two possible routines for realization of stable 2D sumanene monolayers (named as $\alpha$ phase and $\beta$ phase), and $\alpha$-sumanene monolayer can be experimentally synthesized with chemical vapor deposition using $C_{21}H_{12}$ as a precursor. Small binding energies on Au(111) surface (e.g. $-37.86$ eV/Å$^2$ for $\alpha$ phase) signify the possibility of their peel-off after grown on the noble metal substrate. Importantly, our GW plus Bethe-Salpeter equation calculations demonstrate both monolayers have moderate band gaps (1.94 eV for $\alpha$) and ultrahigh carrier mobilities ($3.4 \times 10^4$ cm$^2$V$^{-1}$s$^{-1}$ for $\alpha$). In particular, $\alpha$-sumanene monolayer possesses a strong exciton binding energy of 0.73 eV, suggesting potential applications in optics.

**Keywords:** 2D lattice, Dirac cone, Flat band, Exciton, Sumanene

The relationship between lattice structures and electronic structures of materials is an important topic in condensed matter physics because crystal systems with a similar lattice structure generally tend to share certain key properties. For example, two-dimensional (2D) honeycomb lattice of graphene introduces Dirac cones with nearly zero effective mass, which enables potential electrical[1-2] and optical[3-4] applications because of the high carrier mobility[5] and ballistic transport properties[6]. Apart from graphene, many other honeycomb-lattice materials, such as silicene, germanene, stanene,[7-9] and even monolayers consisting of heavy metal atoms arranged in a honeycomb lattice on substrates[10], also exhibit Dirac cones (Fig. 1a) and topological quantum states.

Recently, Kagome lattice constructed by combining triangle and honeycomb lattices through corner-sharing triangles[11] has attracted intense focus. Due to the geometric frustration of the Kagome lattice, there exist local excited states that lead to flat bands[12]. Meanwhile, the Kagome lattice also exhibits other interesting characters, such as a set of Dirac cones[13-16] as shown in Fig. 1b. The Dirac cones produce massless fermions with extremely high carrier mobility[17-22], while the flat bands produce heavy fermions[23-24] that are associated with topologically non-trivial band structures, ferromagnetism[25-26] and superconductivity[27].

Besides traditional lattices, 2D monolayer materials can also be used to construct complex structures by Moire twisting, self-assembly or surface modification[28-29]. Fundamentally, the physical properties of a monolayer graphene or graphene-like 2D material can be tailored by means of band engineering. For example, opening of band gap, introduction of flat bands and realization of superconductivity in graphene have been achieved by various strategies, such as doping[30], applying strain[31-32], cutting[33], rotation[34-35], and molecular self-assembling[36].

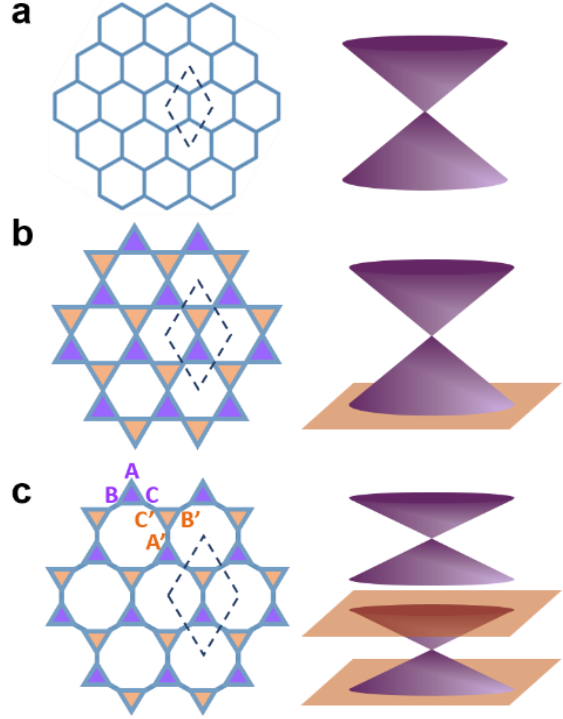

**Fig. 1** (Left panels) Geometrical structures of (**a**) honeycomb, (**b**) Kagome and (**c**) Kagome-analogy sumanene lattices and (Right panels) the schematic diagrams of their low-energy electronic band structure with tight-binding model.

Identification and realization of a new 2D lattice allow the creation of platforms to study its novel physical properties and explore its potential applications. Here, we propose an unprecedented 2D lattice, which is composed of two sets of triangles with bonds linking between each angle point, as shown in Fig. 1c. There are six lattice sites (A, B, C, A', B', and C') in the unit cell. The tight-binding model Hamiltonian for this 2D Kagome-analogy lattice is written as

$$H = \begin{bmatrix} E_0 & p_1 & p_3 & p_4 & 0 & 0 \\ p_1^* & E_0 & p_2 & 0 & p_5 & 0 \\ p_3^* & p_2^* & E_0 & 0 & 0 & p_6 \\ p_4^* & 0 & 0 & E_0 & p_1^* & p_3^* \\ 0 & p_5^* & 0 & p_1 & E_0 & p_2^* \\ 0 & 0 & p_6^* & p_3 & p_2 & E_0 \end{bmatrix}, \quad (1)$$

where $E_0$ is the site energy, $p_1 = te^{\frac{1}{2}i(k_x+\sqrt{3}k_y)a_1}$, $p_2 = te^{-ik_x a_1}$, $p_3 = te^{\frac{1}{2}i(-k_x+\sqrt{3}k_y)a_1}$, $p_4 = t'e^{-ik_y a_2}$, $p_5 = t'e^{\frac{1}{2}i(k_y+\sqrt{3}k_x)a_2}$, and $p_6 = t'e^{\frac{1}{2}i(k_y-\sqrt{3}k_x)a_2}$ with $t$ and $t'$ denoting the hopping parameters, $a_1$ and

$a_2$ being the interatomic distances. Interestingly, the above tight-binding model predicts that this 2D Kagome-analogy sumanene lattice may be a semiconductor with a suitable band gap. Moreover, the band structures of Hamiltonian given by Eq. (1) contain two Dirac cones at K points and two flat bands in the valance bands, which are schematically shown in the right panel of Fig. 1c. Additionally, one flat band gap is degenerate at the valance band maximum (VBM). Certainly, this new 2D material with unique band structure provides a fascinating platform to explore its rich physical characteristics and potential applications in electronic devices. It is thus imperative to realize such exotic band structure in real 2D materials.

However, it is a grand challenge to bring the metaphysics to reality since one must find suitable molecule/motif to fill the proposed lattice and yet comply with the tight-binding model. Fortunately, the ultra-stable $C_{21}$ cluster with $C_{3v}$ symmetry, which has been explored previously, may be a good choice for realizing the low-energy tight-binding Hamiltonian in Eq. (1). Note that $C_{21}$ is a magic cluster having a population of over 90% during the growth of graphene under certain conditions[37-41]. Furthermore, $C_{21}$ cluster is the core of a known hydrocarbon molecule, namely, sumanene ($C_{21}H_{12}$), as shown in Fig. 2a[42]. Hirao et al.[43] even reported that a $C_{21}H_{12}$ molecular monolayer could be achieved by self-assembling on Ag(111) surface. Beyond, they also proposed that the peripheral H atoms of sumanene can be detached, resulting in dimerization of $C_{21}$. As further derivative of dimerization and $C_{21}H_{12}$ molecular monolayer on a certain substrate, dehydrogenated $C_{21}$ can bond with each other and form a unique lattice, named as sumanene monolayer herein. Such derivative should be experimentally realized by using the bottom-up approaches via molecular self-assembly, which is a promising method to fabricate nanomaterials[44-47]. For example, atomically precise zigzag-edged graphene nanoribbons were obtained by molecular self-assembly of certain molecules like U-shaped dibenzo[a,j]anthracene monomer 1 with halogen functions $R_1$ = Br [48].

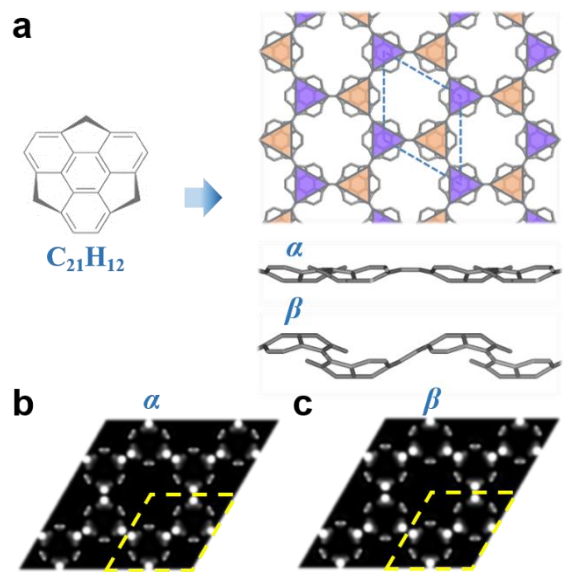

**Fig. 2 (a)** Structure of sumanene molecule ($C_{21}H_{12}$)[42] and the top and side views of $\alpha$-$C_{21}$ and $\beta$-$C_{21}$ monolayers. Simulated STM images of **(b)** $\alpha$ and **(c)** $\beta$ at 0.5 V bias. Unit cells are indicated by the yellow dashed line.

As shown in Fig. 2(a), $C_{21}$ contains three five-membered rings of carbon atoms with included angle of 120°. The carbon atoms at the outer vertices of the five-membered rings can be regarded as lattice points A, B, C, and A', B', and C' in the unique lattice of Fig. 1c. Actually, $C_{21}$ cluster has a bowl structure; therefore, there are two probable configurations of sumanene monolayer, which are named as $\alpha$ and $\beta$ with different arrangements of $C_{21}$, as shown in Fig. 2a. In $\alpha$-type configuration, all $C_{21}$ bowls point up, while $C_{21}$ bowls in $\beta$-type point up and down alternatively. Upon relaxation, the lattice parameters of sumanene monolayer are 13.81 Å ($\alpha$) and 13.16 Å ($\beta$), respectively. The computed phonon dispersions (Fig. S1) indicate that both $\alpha$ and $\beta$ phase of sumanene monolayers are dynamically stable without any imaginary mode.

To guide experimental identification, we simulated scanning tunneling microscope (STM) images of $\alpha$ and $\beta$ at 0.5 V bias, which are shown in Fig. 2b, c. Clearly, the bright spots are mainly located at the three five-membered rings of $C_{21}$, which indeed resemble the unique lattice of Fig. 1c. Fig. S2 shows the band structures of $\alpha$ and $\beta$ sumanene monolayers from density functional theory (DFT) calculation using PBE functional, and the DFT results generally resemble the characteristics of the tight-binding model (Fig. 1c), i.e., there are two Dirac cones (one in the conduction band, and the other in the valance band) and two flat bands (one being degenerated in VBM, and another one located just at the bottom of Dirac cone). Both α and β sumanene monolayers are direct band gap semiconductors with VBM and conduction band minimum (CBM) located at Γ point. At PBE level, the band gaps are 0.58 eV for α phase and 0.07 eV for β phase, respectively.

Generally speaking, 2D materials have weaker dielectric screening than that of bulk counterparts and would result in more prominent excitonic effects, which provide new opportunities for designing new energy conversion and optoelectronic devices[49-51]. Since excitonic effects strongly depend on the accuracy in band structure calculations, here we employed GW approximation to obtain the quasi-particle (QP) band structures of both sumanene monolayers. The GW band gaps for α and β phase of sumanene lattice are 1.94 eV (Fig. 3a) and 0.53 eV (Fig. 3b), respectively, both of which are much larger the PBE values.

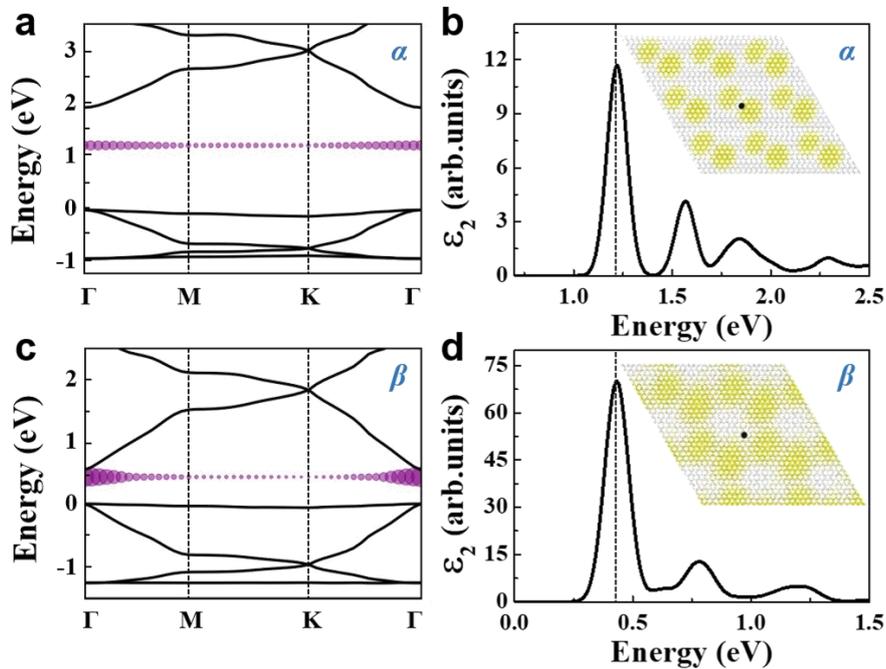

**Fig. 3** Calculated QP band structure (left panels, **a** and **c**) and absorption spectra (right panels, **b** and **d**) of α and β. The dots show the *k*-space wave functions for the lowest-lying exciton (purple). The energy at VBM is set to zero. The real space distribution of excitons (insets in **b** and **d**) at the first peak, and the positions of the holes are indicated by black dots.

Moreover, using GW approach [52-53] plus Bethe-Salpeter[53-54] equation (GW-BSE), the optical absorption spectra of two freestanding sumanene monolayers are calculated, and the results are listed in Table I. It is known that the energy of first peak in the absorption spectrum is regarded as the optical gap. For $α$ phase (Fig. 3c), the first peak is located at 1.21 eV, which corresponds to the lowest-energy exciton with a binding energy ($E_b$) of 0.73 eV. For the $β$ phase (Fig. 3d), the optical gap is 0.4 eV with $E_b$ of 0.1 eV. Our results are consistent with the general trend that exciton binding energy scales linearly with QP band gap in prototypical 2D semiconductors[55]. Remarkably, $α$-sumanene monolayer is found to exhibit a very high exciton binding energy of 0.73 eV, and its prominent excitonic effects can be comparable with that of $h$-$BN$ bulk ($E_b$~0.7 eV)[56], which can be used as room-temperature single photon emitters for quantum technologies[57]. For comparison, $MoS_2$ monolayer, as a typical 2D material and the most studied transition metal dichalcogenides (TMDs), has a quasiparticle bandgap of 2.84 eV and an exciton binding energy of 0.96 eV[58]. Furthermore, excitons in TMDs have been reported to play a dominant role in optical absorption, photoluminescence and spinning valley dynamics even at room temperature[59]. Considering the excellent performance of $h$-$BN$ and $MoS_2$, the sumanene monolayers are anticipated to be potential candidates for various applications in optics and optoelectronics.

$k$-space wave function for the lowest-lying exciton is then analyzed. The lowest-energy exciton wave functions of the two phases are mostly located around Γ point. This indicates that the excitons are of Mott-Wannier type, which should exhibit considerable delocalization in the real space. We also plot the wave function of the exciton in the real space (insets in Fig. 3b and 3d), with the hole fixed at the center. The sizes of the exciton in these two structures are different, but both having $C_{3v}$ symmetrical distribution. The exciton distribution size of $α$ phase is about 5.98 nm and the adjacent region span is 2.39 nm, while the exciton distribution size of $β$ phase is about 7.98 nm and the adjacent region span is 2.28 nm, indeed showing considerable delocalization characteristics in the real space.

**Table I.** Comparison of band gaps and exciton binding energies between α and β phases.

|        | PBE gap | GW gap | Optical gap | Binding energy |
|--------|---------|--------|-------------|----------------|
| **α** (eV) | 0.58 | 1.94 | 1.21 | 0.73 |
| **β** (eV) | 0.07 | 0.53 | 0.43 | 0.10 |

The two sumanene structures are expected to possess superior carrier mobility since they partially retain the properties of graphene. To prove this, we calculated the electron effective mass at CBM and the hole effective mass at VBM (only for the light hole, which is degenerated with flat band for the heave hole). As shown in Table S7, the electron and hole effective masses are remarkably small. For example, the effective mass of electron is $m_e^* = 0.11\ m_0$, while that of hole is $m_h^* = 0.06\ m_0$ in zigzag direction of α-sumanene monolayer. This material also shows isotropic character in the effective mass. Compared with α-sumanene monolayer, both electron and hole effective masses of β-sumanene monolayer are even lower, i.e., only 0.02 $m_0$. Such small electron effective masses are comparable with those along armchair direction in graphene nanoribbons with a ribbon width of 15 nm $(0.05 m_0)^{60}$.

Next, we evaluated their charge mobilities using the deformation potential method[61-62]. The 2D in-plane stiffness ($C_{2D}$), effective mass ($m^*$), deformation potential constant ($E_1$) and carrier mobility ($\mu$) are summarized in Table S7. The electron and hole mobilities of α-sumanene monolayer along the armchair direction of $6.93 \times 10^3$ cm$^2$V$^{-1}$s$^{-1}$ and $4.44 \times 10^3$ cm$^2$V$^{-1}$s$^{-1}$ are comparable to those of phosphorene[63]. Notably, the hole and electron mobilities of β-sumanene monolayer are even higher, i.e., which are of the same order of magnitude as those of graphene ($\sim 10^5$ cm$^2$V$^{-1}$s$^{-1}$)[64].

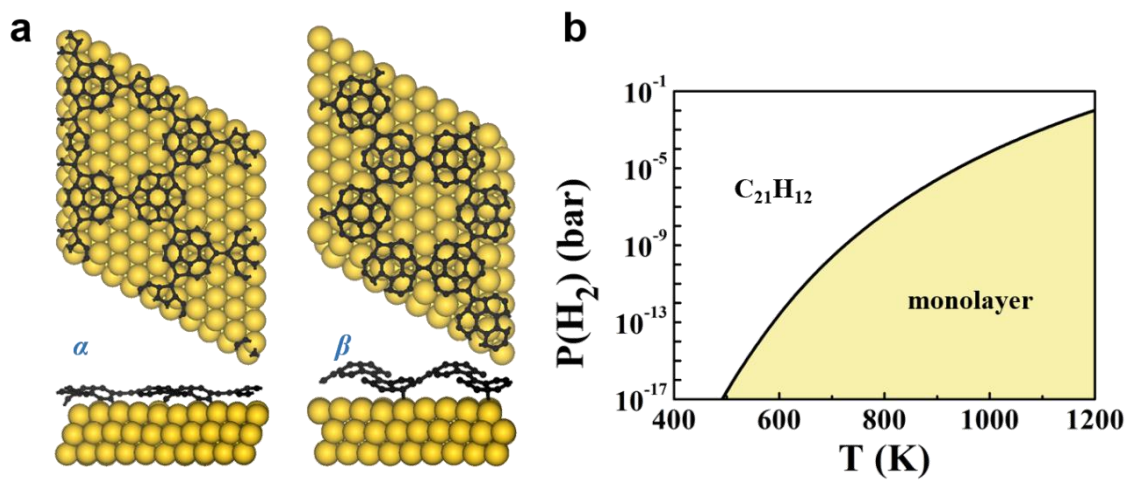

**Fig. 4 (a)** The structures of α and β sumanene monolayers on Au(111) surface, respectively. Black and gold balls represent C and Au atoms, respectively. **(b)** Phase diagram of $C_{21}H_{12}$ molecules and sumanene monolayer under various partial pressures of $H_2$ over temperature range of 400–1200 K on Au(111) surface.

Two-dimensional materials, such as silicene and borophorene[65-68], have been successfully grown on noble metal surfaces owing to their moderate interaction with the substrates, suggesting that sumanene monolayers can be experimentally synthesized with chemical vapor deposition (CVD) on a suitable noble metal surface. More excitingly, both α and β sumanene monolayers are well preserved on Au(111) surface according to our simulations (Fig 4a). The α-sumanene monolayer is 22.34 meV/Å² lower than that of β phase in energy, implying the α-sumanene with larger band gap (1.94 eV) and stronger exciton binding energy (0.73 eV) is easier to form on Au(111) surface.

To further confirm the feasibility of experimental synthesis of these monolayers, we investigated the phase diagram of $C_{21}H_{12}$ molecules and α-sumanene monolayers on Au(111) surface referring to $H_2$ partial pressure and temperature in CVD progress[41]. As shown in Figure 4b, at 500K, the critical value of $P(H_2)$ in the $C_{21}H_{12}$ molecular dehydrogenation into sumanene monolayer is about $3.5 \times 10^{-17}$ bar, excessively lower than the experimental low pressure of ultra-high vacuum (~$10^{-13}$ bars)[38]. Notably, elevating temperature to 900 K (which is the dehydrogenation temperature of $CH_4$), $C_{21}H_{12}$ will transform into more energetically favored α-sumanene on Au(111) surface once P(H$_2$) is lower than 10$^{-6}$ bar (10$^{-1}$ par), which is easy to achieve in CVD experiments such as the graphene growth.[69] Therefore, the α-sumanene monolayer is highly expected to grow on Au(111) surface by using C$_{21}$H$_{12}$ as precursors under low H$_2$ pressure, like the case of graphene[70]. Beyond that, the adsorption energy is only −37.86 meV/Å$^2$ for α-sumanene, lower than 50 meV/Å$^2$. Such a low adsorption energy is comparable with that of graphene on Cu(111) surface,[71] implying freestanding sumanene is likely to be peeled off from substrate and the transfer technology of 2D materials can also be applied to sumanene monolayer.

In summary, we propose a novel 2D lattice possessing two sets of Dirac cones and two flat bands near the Fermi level within the picture of tight-binding model. By filling the lattice sites with C$_{21}$ clusters in different arrangements, two types of monolayer phases, that is, α and β sumanene, are achieved, and both of them are thermodynamically stable. Moreover, we indicate α-sumanene monolayer is highly like to from by using dehydronated C$_{21}$H$_{12}$ as a precursor. Beyond, the moderate interaction energy between the 2D sumanene lattices and Au(111) surface is close to that of graphene on Cu(111) surface, implying that 2D sumanene can be peeled off after grown on Au(111) surface. DFT calculations revealed that both α and β are direct band gap semiconductors with GW band gaps of 1.94 eV and 0.53 eV, respectively. Consistent with the tight-binding model, two sets of Dirac cones and two flat bands exist near the Fermi level for both α and β phases. GW-BSE calculations further reveal that the α-sumanene monolayer has a remarkably strong exciton binding energy of 0.73 eV. Besides, both sumanene monolayers possess ultrahigh carrier mobility in the range of 10$^3$ - 10$^5$ cm$^2$V$^{-1}$s$^{-1}$, which is comparable with that of graphene. Finally, the bowl-structure can be regarded as self-swelling nanobubbles, which may exhibit pseudomagnetic behavior and certainly deserve further study.

**Notes**

The authors declare no competing financial interest.

**Corresponding authors:**
Junfeng Gao*: gaojf@dlut.edu.cn

Zhenguo Fu*: fu_zhenguo@iapcm.ac.cn

**Author Contributions**

The manuscript was written through contributions of all authors. All authors have given approval to the final version of the manuscript.

**Acknowledgements**

This work is supported by the National Natural Science Foundation of China (Grant No. 12074053, 91961204, 12004064), XinLiaoYingCai Project of Liaoning province, China (XLYC1907163, XLYC1905014) and by the Fundamental Research Funds for the Central Universities (DUT21LAB112). We also acknowledge Computers supporting from Shanghai Supercomputer Center, DUT supercomputing center, and Tianhe supercomputer of Tianjin center. This work was partially supported by Singapore NRF-CRP24-2020-0002. Y-WZ acknowledges the support from Singapore A*STAR SERC CRF Award.